\documentclass[12pt]{article}
\usepackage[dvips]{graphicx}
\usepackage{amssymb}
\usepackage{amsmath}
\usepackage{epsfig}
\usepackage{colortbl}
\usepackage{cite}
\usepackage{hyperref}

\numberwithin{equation}{section}
\numberwithin{table}{section}

\def\beq{\begin{equation}}
\def\eeq{\end{equation}}
\def\be{\begin{equation}}
\def\ee{\end{equation}}
\def\bea{\begin{eqnarray}}
\def\eea{\end{eqnarray}}

\DeclareMathOperator{\rk}{rk}

\def\omegaC{\omega}

\def\g{{\mathfrak{g}}}

\def\vev#1{{\langle #1\rangle}}

\DeclareRobustCommand{\SkipTocEntry}[4]{}
\RequirePackage{color}

\def\vev#1{{\langle #1\rangle}}

\newcommand{\cT}{\mathcal{T}}

\newcommand{\cD}{\mathcal{D}}
\newcommand{\cL}{\mathcal{L}}

\newcommand{\cK}{\mathcal{K}}
\newcommand{\cM}{\mathcal{M}}
\newcommand{\cN}{\mathcal{N}}

\newcommand{\cG}{\mathcal{G}}
\newcommand{\cA}{\mathcal{A}}
\newcommand{\cH}{\mathcal{H}}
\newcommand{\cB}{\mathcal{B}}

\newcommand{\cU}{\mathcal{U}}

\newcommand{\bbR}{\mathbb{R}}

\def\AdS{\textrm{AdS}}

\begin{document}
\begin{titlepage}
\begin{center}
\rightline{\small ZMP-HH/16-01}

\vskip 1cm

{\Large \bf Moduli spaces of $\AdS_{5}$ vacua in $\cN=2$ supergravity}
\vskip 1.2cm

{\bf  Jan Louis 
and Constantin Muranaka}

\vskip 0.8cm

{\em Fachbereich Physik der Universit\"at Hamburg, Luruper Chaussee 149, 22761 Hamburg, Germany}
\vskip 0.2cm

and 

\vskip 0.2cm

{\em Zentrum f\"ur Mathematische Physik,
Universit\"at Hamburg,\\
Bundesstrasse 55, D-20146 Hamburg, Germany}
\vskip 0.3cm

\vskip 0.3cm

{\tt jan.louis@desy.de, constantin.muranaka@desy.de}

\end{center}

\vskip 2cm

\begin{center} {\bf ABSTRACT } \end{center}

\noindent
We determine the conditions for  maximally supersymmetric $\AdS_{5}$ vacua of five-dimensional
gauged $\cN=2$ supergravity coupled to vector-, tensor- and
hypermultiplets charged under an arbitrary gauge group.
In particular, we show that the unbroken gauge group of the $\AdS_{5}$ vacua has
to contain an $U(1)_R$-factor. Moreover we prove that 
the scalar deformations which preserve all supercharges
form a K\"ahler submanifold of the ambient quaternionic K\"ahler
manifold spanned by the scalars in the hypermultiplets.

\vfill

January 2016

\end{titlepage}


\tableofcontents


\section{Introduction}

Anti-de Sitter (AdS) backgrounds of supergravity are an essential part of the 
AdS/CFT correspondence \cite{Maldacena:1997re} and  have been studied
in recent years from varying perspectives. On the one hand they can be constructed as compactifications of higher-dimensional supergravities as is the natural
set up in the AdS/CFT correspondence.\footnote{See \cite{Kehagias:1998gn,Morrison:1998cs} for earlier work and 
e.g.\ \cite{Polchinski:2010hw} and references therein for a more recent review.} Alternatively, one can investigate and, if possible, 
classify their appearance directly in a given supergravity without relating it
to any compactification.

For a given AdS background it is also of interest to study its properties
and in particular its moduli space $\cM$, i.e.\ the subspace of the scalar field space
that is spanned by flat directions of the AdS background.
This moduli space has been heavily investigated  
in Minkowskian backgrounds of string theory as it prominently appears
in its low energy effective theory.
For AdS backgrounds much less is known about $\cM$, partly because the defining equations are more involved and furthermore quantum corrections contribute unprotected.


In \cite{deAlwis:2013jaa,Louis:2014gxa} supersymmetric $\AdS_{4}$ vacua 
and their classical supersymmetric moduli spaces
were studied  in four-dimensional ($d=4$) supergravities 
with $\cN=1,2,4$ supersymmetry
without considering their relation to higher-dimensional theories.\footnote{Throughout this paper we only consider $\AdS$ backgrounds that preserve all supercharges of a given supergravity and furthermore only consider the subspace of the moduli space that preserves all these supercharges. This is what we mean by supersymmetric $\AdS$ backgrounds and supersymmetric moduli spaces.}
For $\cN=1$ it was found that  the supersymmetric moduli space is at best a real submanifold of the original K\"ahler field space.
Similarly, for $\cN=2$ the supersymmetric moduli space 
is at best a product of a real manifold times a K\"ahler manifold
while  $\cN=4$ $\AdS$ backgrounds have no supersymmetric moduli space.
 This analysis was repeated for $\AdS_{5}$ vacua in  $d=5$ gauged supergravity
with 16 supercharges ($\cN=4$) in \cite{Louis:2015dca} and for $\AdS_{7}$ vacua in  $d=7$ gauged supergravity with 16~supercharges in \cite{Louis:2015mka}. For the $d=5,\, \cN=4$ theories it was shown that the supersymmetric moduli space is
the coset $\cM=SU(1,m)/(U(1)\times SU(m))$ while in $d=7$ it was proven that again 
no supersymmetric moduli space exists.

In this paper we focus on supersymmetric $\AdS_{5}$ vacua in $d=5$ gauged 
supergravities with eight supercharges ($\cN=2$)
coupled to an arbitrary number of vector-, tensor- and hypermultiplets. 
A related analysis was carried out in \cite{Tachikawa:2005tq}
for the coupling of Abelian vector multiplets and hypermultiplets.
We confirm the results of \cite{Tachikawa:2005tq} and generalize 
the analysis by including tensor multiplets and
non-Abelian vector multiplets. 
In particular, we show that also in this more general case 
the unbroken gauge group has to 
 be of the form $H\times U(1)_{R}$
where the $U(1)_R$-factor is gauged by the graviphoton.
This specifically forbids unbroken semisimple gauge groups in AdS
backgrounds.

In a second step
we study the supersymmetric moduli space $\cM$ 
of the previously obtained  $\AdS_{5}$ backgrounds
and show that it necessarily is a K\"ahler submanifold of the quaternionic scalar field space $\cT_H$ spanned by all scalars in the hypermultiplets.\footnote{This result was also obtained in  \cite{Tachikawa:2005tq}. Our results is more general as we include tensor multiplets and non-Abelian vector multiplet in the analysis.}
This is indeed consistent with  the AdS/CFT correspondence where the 
moduli space $\cM$ is mapped to the conformal manifold of the dual 
superconformal field theory (SCFT). For the gauged supergravities considered here
the dual theories are $d=4,\, \cN=1$ SCFTs.
In \cite{Asnin:2009xx} it was indeed shown that 
the conformal manifold of these SCFTs  is a K\"ahler manifold. 

The organization of this paper is as follows. In section \ref{sec:sugra} we briefly review gauged $\cN=2$ supergravities in five dimensions. This will then be used to study the conditions for the existence of supersymmetric $\AdS_{5}$ vacua and determine some of their properties in section~\ref{sec:vacua}. Finally, in section \ref{sec:moduli} we compute the conditions on the moduli space of these vacua and show that it is a K\"ahler manifold.


\section{Gauged $\cN=2$ supergravity in five dimensions}\label{sec:sugra}

To begin with let us review five-dimensional gauged $\cN=2$ supergravity following  \cite{Gunaydin:2000xk,Bergshoeff:2002qk,Bergshoeff:2004kh}.\footnote{Ref.~\cite{Bergshoeff:2004kh}
constructed the most general version of five-dimensional gauged $\cN=2$ supergravity.} The theory consists of the gravity multiplet with field content
\begin{equation}
\{g_{\mu\nu}, \Psi_{\mu}^{\cA}, A_{\mu}^{0}\}\ , \quad \mu,\nu=0,...,4\ ,\quad 
\cA=1,2\ ,
\end{equation}
where $g_{\mu\nu}$ is the metric of space-time, $\Psi_{\mu}^{\cA}$ is
an $SU(2)_{R}$-doublet of symplectic Majorana gravitini and $A_{\mu}^{0}$ is the graviphoton. In
this paper we consider theories that additionally contain $n_{V}$
vector multiplets, $n_{H}$ hypermultiplets and $n_{T}$ tensor
multiplets. A vector multiplet $\{A_{\mu}, \lambda^{\cA}, \phi\}$
transforms in the adjoint representation of the gauge group $G$ and contains a vector $A_{\mu}$, a doublet of gauginos $\lambda^{\cA}$ and a real scalar~$\phi$. In $d=5$ a vector is Poincar\'e dual  to an antisymmetric 
tensor field $B_{\mu\nu}$ which carry an arbitrary representation of $G$.  This gives rise to tensor multiplets which have the same field content as vector multiplets, but with a two-form instead of a vector. Since vector- and tensor multiplets mix in the Lagrangian, we label their scalars $\phi^{i}$ by the same index $i,j=1,...,n_{V}+n_{T}$. Moreover, we label the vector fields (including the graviphoton) by $I,J=0,1,...,n_{V}$, the tensor fields by $M,N=n_{V}+1,...,n_{V}+n_{T}$ and also introduce a combined index $\tilde{I}=(I,M)$. Finally, the $n_{H}$ hypermultiplets
\begin{equation}
\{q^{u}, \zeta^{\alpha}\}, \quad u=1,2,...,4n_{H}\ , \quad \alpha=1,2,...,2n_{H}\ , 
\end{equation}
contain $4n_{H}$ real scalars $q^{u}$ and $2n_{H}$ hyperini $\zeta^{\alpha}$.

The bosonic Lagrangian of $\cN=2$ gauged supergravity in five dimensions reads\footnote{
Note that we set the gravitational constant $\kappa=1$ in this paper.}
\cite{Bergshoeff:2004kh}
\begin{equation}\label{eq:Lagrangian}
\begin{aligned}
e^{-1}\cL&=\tfrac{1}{2}R
-\tfrac{1}{4}a_{\tilde{I}\tilde{J}}H^{\tilde{I}}_{\mu\nu}H^{\tilde{J}\mu\nu}-\tfrac{1}{2}g_{ij}\cD_{\mu}\phi^{i}\cD^{\mu}\phi^{j}-\tfrac{1}{2}G_{uv}\cD_{\mu}q^{u}\cD^{\mu}q^{v}-g^{2}V(\phi,q)\\
&+\tfrac{1}{16g}e^{-1}\epsilon^{\mu\nu\rho\sigma\tau}\Omega_{MN}B^{M}_{\mu\nu}\left(\partial_{\rho}B^{N}_{\sigma\tau}+2gt_{IJ}^{N}A_{\rho}^{I}F_{\sigma\tau}^{J}+gt_{IP}^{N}A_{\rho}^{I}B_{\sigma\tau}^{P}\right)\\
&+\tfrac{1}{12}\sqrt{\tfrac{2}{3}}e^{-1}\epsilon^{\mu\nu\rho\sigma\tau}C_{IJK}A_{\mu}^{I}\left[F_{\nu\rho}^{J}F_{\sigma\tau}+f_{FG}^{J}A_{\nu}^{F}A_{\rho}^{G}\left(-\tfrac{1}{2}F_{\sigma\tau}^{K}+\tfrac{g^{2}}{10}f_{HL}^{K}A_{\sigma}^{H}A_{\tau}^{L}\right)\right]\\
&-\tfrac{1}{8}e^{-1}\epsilon^{\mu\nu\rho\sigma\tau}\Omega_{MN}t_{IK}^{M}t_{FG}^{N}A_{\mu}^{I}A_{\nu}^{F}A_{\rho}^{G}\left(-\tfrac{g}{2}F_{\sigma\tau}^{K}+\tfrac{g^{2}}{10}f_{HL}^{K}A_{\sigma}^{H}A_{\tau}^{L}\right)
\ .
\end{aligned}
\end{equation}
In the rest of this section we recall the various ingredients which
enter this Lagrangian.
First of all $H^{\tilde{I}}_{\mu\nu}=(F_{\mu\nu}^{I}, B_{\mu\nu}^{M})$
where
$F_{\mu\nu}^{I}=2\partial_{[\mu}A_{\nu]}^{I}+gf_{JK}^{I}A^{J}_{\mu}A^{K}_{\nu}$
are the field strengths with $g$ being the gauge coupling constant.
The scalar fields in $\cL$ can be interpreted as coordinate charts from spacetime $M_{5}$ to a target space $\cT$,
\begin{equation}\label{eq:target space}
\phi^{i} \otimes q^{u}: M_{5} \longrightarrow \cT.
\end{equation}
Locally $\cT$ is a product $\cT_{VT} \times \cT_{H}$ where the first
factor is a projective special real manifold $(\cT_{VT}, g)$ of
dimension $n_{V}+n_{T}$. It is constructed as a hypersurface in an $(n_{V}+n_{T}+1)$-dimensional real manifold $\cH$ with local coordinates $h^{\tilde{I}}$. This hypersurface is defined by 
\begin{equation}\label{eq:polynomial}
P(h^{\tilde{I}}(\phi))=C_{\tilde{I}\tilde{J}\tilde{K}}h^{\tilde{I}}h^{\tilde{J}}h^{\tilde{K}}=1,
\end{equation}
where $P(h^{\tilde{I}}(\phi))$ is a cubic homogeneous polynomial with $C_{\tilde{I}\tilde{J}\tilde{K}}$ constant and completely symmetric. Thus $\cT_{VT}=\{P=1\}\subset \cH$. 

The generalized gauge couplings in \eqref{eq:Lagrangian} correspond to a positive metric on the ambient space $\cH$, given by
\begin{equation}\label{adef}
a_{\tilde{I}\tilde{J}}:=-2C_{\tilde{I}\tilde{J}\tilde{K}}h^{\tilde{K}}+3h_{\tilde{I}}h_{\tilde{J}}\ ,
\end{equation}
where
\begin{equation}\label{eq:hlower}
 h_{\tilde{I}}= C_{\tilde{I}\tilde{J}\tilde{K}}h^{\tilde{J}}h^{\tilde{K}}\ . 
\end{equation}
The pullback metric $g_{ij}$ is the (positive) metric on the hypersurface 
$\cT_{VT}$ and is  given by
\begin{equation}\label{gpull}
g_{ij}:=h_{i}^{\tilde{I}}h_{j}^{\tilde{J}}a_{\tilde{I}\tilde{J}}\ ,
\end{equation}
where
\begin{equation}\label{eq:hder}
h_{i}^{\tilde{I}}:=-\sqrt{\tfrac{3}{2}}\,\partial_{i}h^{\tilde{I}}(\phi)\
.
\end{equation}
These quantities satisfy (see Appendix C in \cite{Bergshoeff:2004kh} for more details)
\begin{equation}
h^{\tilde{I}}h_{\tilde{I}}=1\ ,\qquad
h_{\tilde{I}}h_{i}^{\tilde{I}}=0\ ,\qquad
h_{\tilde{I}}h_{\tilde{J}}+h_{\tilde{I}}^{i}h_{\tilde{J}i}=a_{\tilde{I}\tilde{J}} \ ,
\label{eq:hmetric}
\end{equation}
where we raise and lower indices with the appropriate metrics $a_{\tilde{I}\tilde{J}}$ or $g_{ij}$ respectively.
The metric $g_{ij}$ induces a covariant derivative which acts on the $h^{\tilde{I}}_{i}$ via 
\begin{equation}\label{eq:covderh}
\nabla_{i}h^{\tilde{I}}_{j}=-\sqrt{\tfrac{2}{3}}\, (h^{\tilde{I}}g_{ij}+T_{ijk}h^{\tilde{I}k})\ ,
\end{equation}
where $T_{ijk}:=C_{\tilde{I}\tilde{J}\tilde{K}}h_{i}^{\tilde{I}}h_{j}^{\tilde{J}}h_{k}^{\tilde{K}}$ is a completely symmetric tensor. 

The second factor of $\cT$ in (\ref{eq:target space}) is a quaternionic K\"ahler manifold $(\cT_{H},G, Q)$ of real dimension $4n_{H}$ (see \cite{Andrianopoli:1996cm} for a more extensive introduction). Here $G_{uv}$ is a Riemannian metric and $Q$ denotes a $\nabla^{G}$ invariant rank three subbundle $Q\subset \text{End} (T\cT_H)$ that is locally spanned by a triplet $J^{n}$, $n=1,2,3$ of almost complex structures which satisfy $J^{1}J^{2}=J^{3}$ and $(J^{n})^{2}=-\text{Id}$. Moreover the metric $G_{uv}$ is hermitian with respect to all three $J^{n}$ and one defines the associated triplet of two-forms $\omegaC^{n}_{uv}:=G_{uw}(J^{n})^{w}_{v}$. In contrast to the K\"ahlerian case, the almost complex structures are not parallel but the Levi-Civita connection $\nabla^{G}$ of $G$ rotates the endomorphisms inside $Q$, i.e. 
\begin{equation}\label{nabladef}
\nabla J^{n}:=\nabla^{G}J^{n}-\epsilon^{npq}\theta^{p}J^{q}=0\ .
\end{equation}
Note that $\nabla$ differs from $\nabla^{G}$ by an
$SU(2)$-connection with connection one-forms $\theta^{p}$.
For later use let us note that the metric $G_{uv}$ can be expressed in terms of vielbeins $\cU^{\alpha\cA}_{u}$ as
\begin{equation}
G_{uv}= C_{\alpha\beta}\epsilon_{\cA\cB}\cU^{\alpha\cA}_{u}\cU^{\beta\cB}_{v}\ ,
\end{equation}
where $C_{\alpha\beta}$ denotes the flat metric on $Sp(2n_{H},\bbR)$
and the $SU(2)$-indices $\cA,\cB$ are raised and lowered with $\epsilon_{\cA\cB}$. 

The gauge group $G$ is specified by the generators $t_{I}$ of its Lie algebra $\mathfrak{g}$ and the structure constants $f_{IJ}^{K}$,
\begin{equation}
[t_{I},t_{J}]=-f_{IJ}^{K}t_{K}\ .
\end{equation}
The vector fields  transform in the adjoint representation of the gauge group, i.e.\ $t_{IJ}^{K}=f_{IJ}^{K}$ while the tensor fields
can carry an arbitrary representation.
The most general representation for $n_{V}$ vector multiplets and $n_{T}$ tensor multiplets has been found in \cite{Bergshoeff:2002qk} and is given by
\begin{equation}\label{trep}
 t_{I\tilde{J}}^{\tilde{K}}=
\begin{pmatrix}
f_{IJ}^{K} & t_{IJ}^{N}\\
0 & t_{IM}^{N}\\
\end{pmatrix}.
\end{equation}
We see that the block matrix $t_{IJ}^{N}$ mixes vector- and tensor
fields. However the $t_{IJ}^{N}$ are only nonzero if the chosen
representation of the gauge group is not completely reducible. This
never occurs for compact gauge groups but there exist non-compact
gauge groups containing an Abelian ideal that admit representations
of this type, see
\cite{Bergshoeff:2002qk}. There it is also shown that the construction
of a generalized Chern-Simons term in the action for vector- and
tensor multiplets requires the existence of an invertible and
antisymmetric matrix $\Omega_{MN}$. In particular, the $t_{I\tilde J}^{N}$
are of the form 
\begin{equation}\label{eq:Omega}
t_{I\tilde{J}}^{N}=C_{I\tilde{J}P}\Omega^{PN}\ .
\end{equation}

The gauge group is realized on the scalar fields via the action of
Killing vectors $\xi_{I}$ for the vector- and tensor multiplets and
$k_{I}$ for the hypermultiplets that satisfy the Lie
algebra~$\g$~of~$G$, 
\begin{equation}\label{Killingc}
\begin{aligned}
{}[\xi_{I},\xi_{J}]^{i}&:=\xi_I^j\partial_j \xi^i_J-\xi_J^j\partial_j \xi^i_I=
-f_{IJ}^{K}\, \xi_{K}^{i}\ ,\\
[k_{I},k_{J}]^{u}&:=k_I^v\partial_v k_J^u-k_J^v\partial_v k_I^u=
-f_{IJ}^{K}\,k_{K}^{u}\ .
\end{aligned}
\end{equation}
In the case of the projective special real manifold, one can obtain an explicit expression for the Killing vectors $\xi_{I}^{i}$ given by \cite{Bergshoeff:2004kh}
\begin{equation}\label{eq:VTkilling}
\xi_{I}^{i}:= -\sqrt{\tfrac{3}{2}}\,t_{I\tilde{J}}^{\tilde{K}}h^{\tilde{J}}h^{i}_{\tilde{K}}=-\sqrt{\tfrac{3}{2}}\,t_{I\tilde{J}}^{\tilde{K}}h^{\tilde{J}i}h_{\tilde{K}}\ .
\end{equation}
The second equality is due to the fact that \cite{Gunaydin:1984ak}
\begin{equation}\label{eq:representation0}
t_{I\tilde{J}}^{\tilde{K}}\,h^{\tilde{J}}h_{\tilde{K}}= 0\ ,
\end{equation}
and thus 
\begin{equation}
0=\partial_{i}(t_{I\tilde{J}}^{\tilde{K}}h^{\tilde{J}}h_{\tilde{K}}) = t_{I\tilde{J}}^{\tilde{K}}h^{\tilde{J}}\partial_{i}h_{\tilde{K}}+t_{I\tilde{J}}^{\tilde{K}}(\partial_{i}h^{\tilde{J}})h_{\tilde{K}}\ , 
\end{equation}
which implies\footnote{Note that the derivative
  $h_{\tilde{I}i}=\sqrt{\tfrac{3}{2}}\,\partial_{i}h_{\tilde{I}}$
has an additional minus sign  compared to \eqref{eq:hder} which can be
shown by lowering the index with $a_{\tilde{I}\tilde{J}}$ given in \eqref{adef}.}
\begin{equation}\label{eq:representation}
t_{I\tilde{J}}^{\tilde{K}}h^{\tilde{J}}h^{i}_{\tilde{K}}=t_{I\tilde{J}}^{\tilde{K}}h^{\tilde{J}i}h_{\tilde{K}}\ .
\end{equation}

The Killing vectors $k_{I}^u$ on the quaternionic K\"ahler
manifold $\cT_H$ \cite{Andrianopoli:1996cm,Alekseevsky:2001if,Bergshoeff:2002qk} have to be triholomorphic which implies 
\begin{equation}\label{eq:Jinvariance}
\nabla_{u}
k^{I}_{w}(J^{n})_{v}^{w}-(J^{n})_{u}^{w}\nabla_{w}k^{I}_{v}=2\epsilon^{npq}\omegaC^{p}_{uv}\mu^{Iq}\
.
\end{equation}
Here $\mu_{I}^{n}$ is a
triplet of moment maps which also satisfy
\begin{equation}\label{eq:covdermomentmap}
\tfrac{1}{2}\omegaC^{n}_{uv}k_{I}^{v}=-\nabla_{u}\mu_{I}^{n}
\ ,
\end{equation}
and the equivariance condition
\begin{equation}\label{eq:equivariance}
f_{IJ}^{K}\mu_{K}^{n}=\tfrac{1}{2}\omegaC_{uv}^{n}k_{I}^{u}k_{J}^{v}-2\epsilon^{npq}\mu_{Ip}\mu_{Jq}\ .
\end{equation}
Furthermore the covariant derivative of the Killing vectors 
obeys \cite{D'Auria:2001kv,Alekseevsky:2001if}
\begin{equation}\label{eq:covderkilling}
\nabla_{u}k_{Iv} +\nabla_{v}k_{Iu} = 0\ ,\qquad \nabla_{u}k_{Iv} -\nabla_{v}k_{Iu} = \omegaC^{n}_{uv}\mu_{nI}+L_{Iuv} \ ,
\end{equation}
where 
the $L_{Iuv}$ are related to the gaugino mass matrix and commute with
$J^{n}$.
For later use we define
\begin{equation}\label{SLdef}
S^{n}_{Iuv}:={L}_{Iuw}(J^{n})^{w}_{v}\ ,\qquad L_{uv}:=h^{I}L_{Iuv}\
,\qquad S_{uv}^{n}:=h^{I}S_{Iuv}^{n}\ ,
\end{equation}
where the $S^{n}_{Iuv}$ are symmetric in $u,v$ \cite{Alekseevsky:2001if}. 

Before we proceed let us 
note that for $n_{H}=0$, i.e.\ when there are no hypermultiplets,
constant Fayet-Iliopoulos (FI) terms can exist which have to satisfy
the equivariance condition \eqref{eq:equivariance}.  
In this case the first term on the right hand side of
\eqref{eq:equivariance}
vanishes which implies that 
 there
are only two possible solutions \cite{Bergshoeff:2004kh}. 
If the gauge group contains an $SU(2)$-factor, the FI-terms have to be
of the form
\begin{equation}
\mu_{I}^{n}= c e_{I}^{n}\ ,\quad c \in \bbR\ ,
\end{equation}
where the $e_{I}^{n}$ are nonzero constant vectors for $I=1,2,3$ of
the $SU(2)$-factor that satisfy
\begin{equation}
 \epsilon^{mnp}e^{m}_{I}e^{n}_{J}=f_{IJ}^{K}e^{p}_{K}\ .
\end{equation}
 The second solution has  $U(1)$-factors in the gauge group and the constant moment maps are  given by 
\begin{equation}\label{eq:AbelianFI}
 \mu_{I}^{n}=c_{I}e^{n}\ ,\quad c_{I}\in \bbR\ ,
\end{equation}
where $e^{n}$ is a constant $SU(2)$-vector and
$I$  labels  the $U(1)$-factors.  

Finally, the covariant derivatives of the scalars in \eqref{eq:Lagrangian} are given by
\begin{equation}\label{eq:covderivatives} 
\cD_{\mu}\phi^{i} = \partial_{\mu}\phi^{i} + gA_{\mu}^{I}\xi_{I}^{i}(\phi)\ , \qquad \cD_{\mu} q^{u} = \partial_{\mu}q^{u}+gA_{\mu}^{I}k_{I}^{u}(q)\ .
\end{equation}
The scalar potential
\begin{equation}\label{eq:potential}
V=2g_{ij}W^{i\cA\cB}W_{\cA\cB}^{j}+2g_{ij}\cK^{i}\cK^{j}+2N^{\alpha}_{\cA}N_{\alpha}^{\cA}-4S_{\cA\cB}S^{\cA\cB},
\end{equation}
is defined in terms of the couplings\footnote{Note that the $h^{M}$ in
  the direction of the tensor multiplets do not appear
  explicitly. Nevertheless, the couplings can implicitly depend on the
  scalars in the tensor multiplet as they might appear in $h^{I}$
  after solving \eqref{eq:polynomial}.}
\begin{equation}\label{eq:definitions}
\begin{aligned}
S^{\cA\cB}&:=h^{I}\mu_{I}^{n}\sigma_{n}^{\cA\cB}\ ,\qquad
W_{i}^{\cA\cB}:=h^{I}_{i}\mu^{n}_{I}\sigma_{n}^{\cA\cB}\ ,\\
\cK^{i}&:=\tfrac{\sqrt{6}}{4} h^{I}\xi_{I}^{i}\ ,\qquad
N^{\alpha\cA}:=\tfrac{\sqrt{6}}{4} h^{I}k_{I}^{u}\cU_{u}^{\alpha\cA}\ .
\end{aligned}
\end{equation}
Here $\sigma^{n}_{\cA\cB}$ are the Pauli matrices with an index lowered by $\epsilon_{\cA\cB}$, i.e.
\begin{equation}
\sigma^{1}_{\cA\cB}= \begin{pmatrix}  1 & 0 \\  0 & -1  \end{pmatrix}\
,\quad
\sigma^{2}_{\cA\cB}= \begin{pmatrix}  -i & 0 \\  0 & -i \end{pmatrix}\
, \quad
\sigma^{3}_{\cA\cB}= \begin{pmatrix}  0 & -1 \\  -1 & 0 \end{pmatrix}\
.
\end{equation}
As usual the  couplings \eqref{eq:definitions}
are related to the
scalar parts of the supersymmetry variations of the fermions via
\begin{equation}\label{susytrans}
\begin{aligned}
\delta_{\epsilon}\psi_{\mu}^{\cA}&=D_{\mu}\epsilon^{\cA}-\tfrac{ig}{\sqrt{6}}S^{\cA\cB}\gamma_{\mu}\epsilon_{\cB}+...\ , \\
\delta_{\epsilon}\lambda^{i\cA}&=g\cK^{i}\epsilon^{\cA}-gW^{i\cA\cB}\epsilon_{\cB}+...\ ,\\
\delta_{\epsilon}\zeta^{\alpha}&=gN_{\cA}^{\alpha}\epsilon^{\cA}+...\ .
\end{aligned}
\end{equation}
Here $\epsilon^{\cA}$ denote the supersymmetry parameters. This concludes our review of $d=5$ supergravity and we now turn to its possible supersymmetric $\AdS$ backgrounds.


\section{Supersymmetric $\AdS_{5}$ vacua}\label{sec:vacua}

In this section we determine the conditions that lead to
 $\AdS_{5}$ vacua which preserve all eight supercharges.
This requires the vanishing of all fermionic 
supersymmetry transformations, i.e.
\begin{equation}
\vev{\delta_{\epsilon}\psi_{\mu}^{\cA}}=\vev{\delta_{\epsilon}\lambda^{i\cA}}=\vev{\delta_{\epsilon}\zeta^{\alpha}}=0 \ ,
\end{equation}
where $\vev{\ }$ denotes the value of a quantity
evaluated in the background. Using the fact that $W^{i\cA\cB}$ and $\cK^{i}$ are linearly
independent \cite{Gunaydin:2000xk} and \eqref{susytrans}, this implies the following four conditions,
\begin{equation}\label{eq:conditions}
\vev{W_{i}^{\cA\cB}}=0\ , \quad \vev{S_{\cA \cB}}\,\epsilon^{\cB}=\Lambda U_{\cA\cB}\,\epsilon^{\cB}\ ,\quad \vev{N^{\alpha\cA}}=0\ ,\quad \vev{\cK^{i}}=0\ .
\end{equation}
Here $\Lambda \in \bbR$ is related to the cosmological constant and
$U_{\cA\cB}=v_{n}\sigma_{\cA\cB}^{n}$  for $v\in S^{2}$ is an $SU(2)$-matrix.
$U_{\cA\cB}$ appears in the Killing spinor equation for $\AdS_{5}$ which reads \cite{Shuster:1999zf}
\begin{equation}
 \vev{D_{\mu}\epsilon_{\cA}}=\tfrac{ia}{2}\, U_{\cA\cB}\,\gamma_{\mu}\epsilon^{\cB}\ , \quad a\in \bbR\ .
\end{equation}
As required for an $\AdS$ vacuum, the conditions \eqref{eq:conditions}
give a negative background value for the scalar potential
$\vev{V(\phi,q)}< 0$ which can be seen from (\ref{eq:potential}).
Using the definitions (\ref{eq:definitions}), we immediately see that 
the four conditions \eqref{eq:conditions} can also be formulated as 
conditions on the moment maps and Killing vectors,
\begin{equation}\label{eq:backgroundmomentmaps}
\vev{h^{I}_{i}\mu^{n}_{I}}=0\ ,\qquad
\vev{h^{I}\mu^{n}_{I}}=\Lambda v^{n}\ ,\qquad
\vev{h^{I}k_{I}^{u}}=0\ , \qquad \vev{h^{I}\xi_{I}^{i}}=0\ .
\end{equation}
 Note that  due to \eqref{eq:polynomial}, \eqref{gpull} we need to have
 $\vev{h^{I}}\neq0$ for some $I$ and $\vev{h^{\tilde I}_i}\neq0$ for every $i$ and some $\tilde I$.\footnote{
In particular this can also hold at the
 origin of the scalar field space $\vev{\phi^i}=0$, i.e.\ for unbroken gauge groups.}

In order to solve \eqref{eq:backgroundmomentmaps} we combine
the first two conditions  as
\begin{equation}\label{eq:momentummaps}
 \vev{\begin{pmatrix}h^{I} \\ h^{I}_{i} \end{pmatrix} \mu_{I}^{n}} = \begin{pmatrix}\Lambda v^{n} \\ 0\end{pmatrix}.
\end{equation}
Let us enlarge these equations to the tensor multiplet indices by introducing $\mu_{\tilde{I}}^{n}$ where we keep in mind that $\mu^{n}_{N}\equiv0$. Then we use the fact that the matrix $(h^{\tilde{I}},h^{\tilde{I}}_{i})$ is invertible in special real geometry (see Appendix C of \cite{Bergshoeff:2004kh}), so we can multiply (\ref{eq:momentummaps}) with $(h^{\tilde{I}},h^{\tilde{I}}_{i})^{-1}$ to obtain a solution for both equations given by
\begin{equation}
\vev{\mu_{\tilde{I}}^{n}}=\Lambda v^{n}\vev{h_{\tilde{I}}}\ .
\end{equation}
Note that this condition is non-trivial since it implies that the moment maps point in the same direction in $SU(2)$-space for all $I$.
Furthermore, using the $SU(2)_{R}$-symmetry we can rotate the vector $v^{n}$
such that $v^{n}=v\delta^{n3}$ and absorb the constant $v\in \bbR$
into $\Lambda$. 
Thus only $\vev{\mu_{I}}:=\vev{\mu_{I}^{3}}\neq 0$, $\forall I$ in the above equation. Since by definition $\vev{\mu_{N}^{n}}= 0$, this implies
\begin{equation}\label{eq:momentmapsvacuum}
\vev{\mu_{I}}=\Lambda \vev{h_{I}}\ , \quad \vev{h_{N}}= 0\ .
\end{equation}
In particular, this means that the first two equations in \eqref{eq:hmetric} hold in the vacuum for only the vector indices, i.e.\
\begin{equation}\label{eq:hmetricvacuum}
\vev{h^{I}h_{I}}=1\ , \quad \vev{h_{I}h^{I}_{i}}=0\ .
\end{equation}
Moreover due to the explicit form of the moment maps in \eqref{eq:momentmapsvacuum}, the equivariance condition \eqref{eq:equivariance} reads in the background
\begin{equation}\label{equivariancevacuum}
 f_{IJ}^{K}\vev{\mu_{K}}=\tfrac{1}{2}\vev{\omegaC^{3}_{uv}k_{I}^{u}k_{J}^{v}}.
\end{equation}

Since \eqref{eq:potential} has to hold in the vacuum, $\vev{h^{I}}\neq 0$ for some $I$ and thus the background necessarily has non-vanishing moment maps due to \eqref{eq:momentmapsvacuum}. This in turn implies that part of the $R$-symmetry is gauged, as can be seen from the covariant derivatives of the fermions which always contain a term of the form $A_{\mu}^{I}\vev{\mu_{I}^{3}}$ \cite{Bergshoeff:2004kh}. More precisely, this combination gauges the $U(1)_{R}\subset SU(2)_{R}$ generated by $\sigma^{3}$. From \eqref{eq:momentmapsvacuum} we infer $A_{\mu}^{I}\vev{\mu_{I}^{3}}=\Lambda A_{\mu}^{I}\vev{h_{I}}$ which can be identified with the graviphoton \cite{Gunaydin:1984ak}.

We now turn to the last two equations in \eqref{eq:backgroundmomentmaps}. Let us first prove that the third equation $\vev{h^{I}k_{I}^{u}}=0$ implies the fourth $\vev{h^{I}\xi_{I}^{i}}=0$. This can be shown by expressing $\vev{\xi_{I}^{i}}$ in terms of $\vev{k_{I}^{u}}$ via the equivariance condition \eqref{equivariancevacuum}. Note that we learn from \eqref{eq:VTkilling} that the background values of the Killing vectors on the manifold $\cT_{VT}$ are given by
\begin{equation}\label{xinA}
 \vev{\xi_{I}^{i}}=
-\sqrt{\tfrac{3}{2}}\,\vev{t_{I\tilde{J}}^{\tilde{K}}h^{\tilde Ji}h_{\tilde K}}
=-\sqrt{\tfrac{3}{2}}\,\vev{f_{IJ}^{K}h^{Ji}h_{K} + t_{IJ}^{N}h^{Ji}h_{N}}
=-\sqrt{\tfrac{3}{2}}\,\vev{f_{IJ}^{K}h^{Ji}h_{K}}
\ ,
\end{equation}
where we used \eqref{trep} and \eqref{eq:momentmapsvacuum}. Inserting \eqref{eq:momentmapsvacuum}, \eqref{equivariancevacuum}  into \eqref{xinA} one indeed computes
\begin{equation}\label{eq:Killingvacuum}
\vev{\xi_{I}^{i}} =
-\sqrt{\tfrac{3}{2}}\tfrac{1}{2\Lambda}\,\vev{h^{J}_{i}\omegaC_{uv}^{3}k_{I}^{u}k_{J}^{v}}\
.
\end{equation}
But then $\vev{h^{I}\xi_{I}^{i}}=0$ is always satisfied if $\vev{h^{I}k_{I}^{u}}=0$. Moreover this shows that $\vev{\xi_{I}^{i}}\neq0$ is only
possible for  $\vev{k^u_I}\neq0$. Note that the reverse is not true in general as can be seen from \eqref{xinA}.
We are thus left with analyzing the third condition in \eqref{eq:backgroundmomentmaps}.

Let us first note that for $n_{H}=0$ there are no Killing vectors ($k_I^u\equiv0$) and the third equation in \eqref{eq:backgroundmomentmaps} is automatically satisfied.
However \eqref{eq:momentmapsvacuum} can nevertheless hold if the constant FI-terms discussed below \eqref{SLdef} are of the form given in \eqref{eq:AbelianFI} and thus only gauge groups with Abelian factors are allowed in this case.

Now we turn to $n_{H}\neq 0$.  Note that then $\vev{h^{I}k_{I}^{u}}=0$ has two possible solutions:
\begin{equation}\begin{aligned}\label{twocases}
i)& \quad \vev{k_{I}^{u}}=0\ ,\quad \textrm{for all}\ I\\
ii)&\quad \vev{k_{I}^{u}}\neq0 \ ,\quad \textrm{for some}\ I \ \textrm{with}\  \vev{h^{I}}\ \textrm{appropriately tuned}.
\end{aligned}\end{equation}
By examining the covariant derivatives (\ref{eq:covderivatives}) of the scalars we see that in the first case there is no gauge symmetry breaking by the hypermultiplets while in the second case $G$ is spontaneously broken. 
Note that not all possible gauge groups can remain unbroken in the vacuum. In fact, for case $i)$ the equivariance condition \eqref{equivariancevacuum} implies
\begin{equation}
 f_{IJ}^{K}\vev{\mu_{K}}=0\ .
\end{equation}
This can only be satisfied if the adjoint representation of $\g$ has a non-trivial zero eigenvector, i.e.\ if the center of $G$ is non-trivial (and continuous).\footnote{For more details on Lie groups and their adjoint representation, see for example \cite{O'Raifeartaigh:1986vq}.} In particular, this holds for all gauge groups with an Abelian factor but all semisimple gauge groups have to be broken in the vacuum.

In the rest of this section we discuss the spontaneous symmetry
breaking for case $ii)$ and the details
of the Higgs mechanism.
Let us first consider the case where only a set of Abelian factors in $G$
is spontaneously broken, i.e.\ $\vev{k^u_I}\neq0$ for $I$ labeling
these Abelian factors.
From \eqref{xinA} we then learn 
$\vev{\xi_{I}^{i}}=0$ and 
thus we only have spontaneous symmetry breaking in the hypermultiplet
sector
and the Goldstone bosons necessarily are recruited out of  these 
hypermultiplets.
Hence the vector multiplet corresponding to a broken Abelian factor in
$G$ becomes massive by ``eating'' an entire hypermultiplet. 
It forms a ``long'' vector multiplet containing the massive vector,
four gauginos and four scalars obeying the AdS mass relations.

Now consider spontaneously broken non-Abelian factors of $G$,
i.e.\ $\vev{k^u_I}\neq0$ for $I$ labeling
these non-Abelian factors.
In this case we learn from \eqref{eq:Killingvacuum} 
that either $\vev{\xi_{I}^{i}}=0$ as before or $\vev{\xi_{I}^{i}}\neq 0$.
However the Higgs mechanism is essentially unchanged compared to the Abelian
case in that entire hypermultiplets are eaten and  all massive vectors
reside in long multiplets.\footnote{Note that short BPS vector
  multiplets which exist in this theory cannot appear since the breaking
  necessarily involves the hypermultiplets.} 

However there always has to exists at least one unbroken generator of
$G$ which commutes with all other unbroken generators, i.e.\ the
unbroken gauge group in the vacuum is always of the form $H\times
U(1)_{R}$. To see this, consider the mass matrix $M_{IJ}$ of the gauge
bosons $A^{I}_{\mu}$. 
Due to \eqref{eq:covderivatives} and \eqref{eq:Killingvacuum}, this is given by
\begin{equation}
 M_{IJ} = \vev{G_{uv}k_{I}^{u}k_{J}^{v}}+\vev{g_{ij}\xi_{I}^{i}\xi_{J}^{j}}=\vev{K_{uv}k^{u}_{I}k^{v}_{J}}\ .
\end{equation}
Here $K_{uv}$ is an invertible matrix which can be given in terms of $G_{uv}$ and $S_{uv}$ defined in \eqref{SLdef} as
\begin{equation}
 K_{uv} = \vev{\left(\tfrac{5}{8}G_{uv}-\tfrac{6}{8\Lambda}S_{uv}\right)}\ .
\end{equation}
Since $\vev{h^{I}k_{I}^{u}}=0$ the mass matrix $M_{IJ}$ has a zero
eigenvector given by $\vev{h^{I}}$, i.e.\ the graviphoton
$\vev{h^{I}}A_{I}^{\mu}$ always remains massless in the vacuum. In the
background the commutator of the corresponding Killing vector $h^{I}k_{I}^{u}$ with any other isometry $k_{J}$ is given by
\begin{equation}
 \vev{[h^{I}k_{I}, k_{J}]^{u}} =
 \vev{h^{I}(k_{I}^{v}\partial_{v}k_{J}^{u}-k_{J}^{v}\partial_{v}k_{I}^{u})}=
 -\vev{h^{I}k_{J}^{v}\partial_{v}k_{I}^{u}}\ .
\end{equation}
This vanishes for $\vev{k_{J}^{u}}=0$ and thus the $R$-symmetry
commutes with every other symmetry generator of the vacuum, i.e.\ the
unbroken gauge group is $H \times U(1)_{R}$. In particular, every
gauge group $G$ which is not of this form has to be broken $G \rightarrow H\times U(1)_{R}$.

Let us close this section with the observation that the number of broken generators is determined by the number of linearly
independent $\vev{k_{I}^{u}}$. This coincides with the number of
Goldstone bosons $n_{G}$. In fact the $\vev{k_{I}^{u}}$ form a basis in the
space of
Goldstone bosons  $\cG$ and we have $\cG=\text{span}_{\bbR}\{\vev{k_{I}^{u}}\}$ with $\text{dim}(\cG) = \rk \vev{k_{I}^{u}} = n_{G}$.

In conclusion, we have shown that the conditions for maximally supersymmetric $\AdS_{5}$ vacua are given by
\begin{equation}
 \vev{\mu_{I}}=\Lambda\, \vev{h_{I}}, \quad \vev{h_{M}}=0, \quad \vev{h^{I}k_{I}^{u}}=\vev{h^{I}\xi_{I}^{i}}=0\ .
\end{equation}
Note that the tensor multiplets enter in the final result only implicitly since the $h^{I}$ and its derivatives are functions of all scalars $\phi^{i}$.
The first equation implies that a $U(1)_{R}$-symmetry is always gauged
by the graviphoton while the last equation shows that the unbroken
gauge group in the vacuum is of the form $H\times U(1)_{R}$.  This
reproduces the result of \cite{Tachikawa:2005tq} that the $U(1)_{R}$
has to be unbroken and gauged in a maximally supersymmetric $\AdS_{5}$
background. In the dual four-dimensional SCFT this $U(1)_{R}$ is
defined by a-maximization. Moreover we discussed that if the gauge
group is spontaneously broken the  massive vector multiplets
are long multiplets. 
Finally, we showed that space of Goldstone bosons is given by
$\cG=\text{span}_{\bbR}\{\vev{k_{I}^{u}}\}$ which will be used in the next section to compute the moduli space $\cM$ of these vacua.


\section{Structure of the moduli space}\label{sec:moduli}

We now turn to the computation of the moduli space $\cM$ of the maximally supersymmetric $\AdS_{5}$ vacua determined in the previous section.
Let us denote by $\cD$ the space of all possible deformations of the
scalar fields $\phi\rightarrow \vev{\phi}+\delta \phi$, $q\rightarrow
\vev{q}+\delta q$ that leave the conditions 
\eqref{eq:backgroundmomentmaps} invariant. However, if the gauge group
is spontaneously broken the corresponding Goldstone bosons are among
these deformations but they should not be counted as moduli. Thus the
moduli space is defined as the space of deformations $\cD$ modulo the space of
Goldstone bosons $\cG$, i.e.\ $\cM=\cD / \cG$. 
In order to determine $\cM$ we vary (\ref{eq:backgroundmomentmaps})
to linear order and characterize the space $\cD$ spanned by $\delta \phi$
and $\delta q$ that are not fixed.\footnote{Since we consider the
  variations of the vacuum equations \eqref{eq:backgroundmomentmaps}
  to first order in the scalar fields, this procedure only gives a
  necessary condition for the moduli space.} We then show that the
Goldstone bosons also satisfy the equations defining $\cD$ and
determine the quotient $\cD / \cG$. 

Let us start by varying the second condition of (\ref{eq:backgroundmomentmaps}). This yields
\begin{equation}
\vev{\delta(h^{I}\mu^{n}_{I})}= \vev{(\partial_{i}h^{I})\,\mu^{n}_{I}}\,\delta\phi^{i}+\vev{h^{I}\nabla_{u}\mu^{n}_{I}}\,\delta q^{u}=-\tfrac{1}{2}\vev{\omegaC_{uv}^{n}h^{I}k_{I}^{v}}\delta q^{u}\equiv 0\ ,
\end{equation}
where we used (\ref{eq:backgroundmomentmaps}) and
(\ref{eq:covdermomentmap}). 
Since this variation vanishes automatically, no conditions are imposed on the scalar field variation.

The variation of the first condition in (\ref{eq:backgroundmomentmaps}) gives
\begin{equation}\label{varone}
\begin{aligned}
\vev{\delta(h_{i}^{I}\mu^{n}_{I})}&=\vev{(\nabla_{j}h_{i}^{I})\,\mu^{n}_{I}}\,\delta\phi^{j}+\vev{h_{i}^{I}\nabla_{u}\mu_{I}^{n}}\,\delta q^{u}\\
&=-\sqrt{\tfrac{2}{3}}\vev{\mu^{n}_{I}(h^{I}g_{ij}+h^{Ik}T_{ijk})}\,\delta \phi^{j}-\tfrac{1}{2}\vev{h^{I}_{i}\omegaC^{n}_{uv}k^{v}_{I}}\,\delta q^{u}\\
&=-\sqrt{\tfrac{2}{3}}\Lambda \delta^{n3} \delta\phi_{i}-\tfrac{1}{2}\vev{h^{I}_{i}\omegaC^{n}_{uv}k^{v}_{I}}\,\delta q^{u}=0\ ,
\end{aligned}
\end{equation}
where in the second step we used (\ref{eq:covderh}), (\ref{eq:covdermomentmap})
while in the third  we used (\ref{eq:backgroundmomentmaps}). 
For $n=1,2$ \eqref{varone} imposes
\begin{equation}
\langle h^{I}_{i}\omegaC_{uv}^{1,2}k^{v}_{I}\rangle\, \delta q^{u} = 0\ , \label{eq:12}
\end{equation}
while 
for $n=3$ the deformations $\delta \phi_{i}$ can be expressed in terms of $\delta q^{u}$ as
\begin{equation} \label{eq:deltaphi}
 \delta \phi_{i} = -\sqrt{\tfrac{3}{2}}\tfrac{1}{2\Lambda}\vev{h_{i}^{I}\omegaC_{uv}^{3}k_{I}^{v}}\, \delta q^{u}\ .
\end{equation}
Thus all deformations $\delta \phi_{i}$ are fixed either to vanish or to be related to $\delta q^{u}$. In other words, the entire space of deformations can be spanned by scalars in the hypermultiplets only, i.e.\ $\cD\subset \cT_{H}$. Note that this is in agreement with \eqref{eq:Killingvacuum} and also $\cG \subset \cT_{H}$.

Finally, we vary the third condition in (\ref{eq:backgroundmomentmaps}) to obtain
\begin{equation}
\vev{\delta(h^{I}k_{Iu})}=\vev{\partial_{i}h^{I}k_{Iu}}\,\delta\phi^{i}+\vev{h^{I}\nabla_{v}k_{Iu}}\,\delta q^{v}=0.
\end{equation}
Inserting \eqref{eq:deltaphi} and using \eqref{eq:hmetric}, (\ref{eq:backgroundmomentmaps}) we find
\begin{equation}\label{eq:Killing1}
\big(\tfrac{1}{2\Lambda}\vev{k^{Iu}\omegaC^{3}_{vw}k_{I}^{w}} + \vev{h^{I}\nabla_{v}k_{I}^{u}}\big)\,\delta q^{v} = 0\ .
\end{equation}
Thus we are left with the two conditions \eqref{eq:12} and
 \eqref{eq:Killing1} whose solutions determine $\cD$. For a generic supergravity we will not solve them here in general. However the conditions alone suffice to prove that the moduli space is a K\"ahler submanifold of $\cT_H$ as we will now show.

As a first step we prove that the Goldstone bosons satisfy \eqref{eq:12} and  \eqref{eq:Killing1}.
We know from section~\ref{sec:vacua} that the Goldstone directions are
of the form $\delta q^{u} = c^I\vev{k_{I}^{u}}$ where  $c^I$ are constants.
Inserted into \eqref{eq:12} we find
\begin{equation}
c^I\vev{h_{i}^{J}\omegaC_{uv}^{1,2}k^{u}_{I}k^{v}_{J}}=2c^I\vev{h_{i}^{J}f_{IJ}^{K}\mu_{K}^{1,2}}
= 0\ ,
\end{equation}
where we used (\ref{equivariancevacuum}) and the fact that $\vev{\mu_{K}^{1,2}}=0$.
To show that the Goldstone bosons also satisfy (\ref{eq:Killing1})
we first  observe that
\begin{equation}\label{eq:killingalgebra2}
 \vev{h^{I}(\nabla_{v}k_{I}^{u})k^{v}_{J}}= \vev{h^{I}(\partial_{v}k_{I}^{u})k_{J}^{v}-h^{I}(\partial_{v}k_{J}^{u})k_{I}^{v}} = -\vev{h^{I}[k_{I},k_{J}]^{u}} = \vev{f_{IJ}^{K}h^{I}k_{K}^{u}}\ ,
\end{equation}
where 
in the first step we used \eqref{eq:backgroundmomentmaps},
added a term which vanishes in the
background
 and then in the second step used \eqref{Killingc}.
In addition  we need to show
\begin{equation}\label{eq:structureconstants}
\vev{f_{IJ}^{K}h^{I}k_{K}^{u}}=\vev{f_{IJ}^{K}h_{K}k^{Iu}}\ .
\end{equation}
Indeed, using \eqref{eq:hmetric} and $\vev{h^{I}k_{I}^{u}}=0$ we find
\begin{equation}
\vev{f_{IJ}^{K}h^{I}k_{K}^{u}}=\vev{f_{IJ}^{K}h^{I}k^{Lu}a_{KL}}=\vev{f_{IJ}^{K}h^{I}k^{Lu}h_{K}^{i}h_{Li}}\ .
\end{equation}
Inserting \eqref{eq:representation} evaluated in the vacuum, i.e.\
$\vev{f_{IJ}^{K}h^{J}h_{K}^{i}}=\vev{f_{IJ}^{K}h^{Ji}h_{K}}$ and using
again \eqref{eq:hmetric}
we obtain 
\begin{equation}
\vev{f_{IJ}^{K}h^{I}k_{K}^{u}}
=\vev{f_{IJ}^{K}h^{Ii}k^{Lu}h_{K}h_{iL}}=\vev{f_{IJ}^{K}h_{K}k^{Lu}\delta^{I}_{L}}=\vev{f_{IJ}^{K}h_{K}k^{Iu}}\ ,
\end{equation}
which proves \eqref{eq:structureconstants} as promised.

Turning back to \eqref{eq:Killing1}, we insert $\delta q^{u}= c^{I}
\vev{k_{I}^{u}}$ and use \eqref{equivariancevacuum} and \eqref{eq:killingalgebra2}
to arrive at
\begin{equation}\label{GBint}
\tfrac{1}{2\Lambda}c^{I}\vev{k^{Ju}\omegaC^{3}_{vw}k_{J}^{w}k_{I}^{v}}+c^{I}\vev{h^{J}\nabla_{v}k_{J}^{u}k_{I}^{v}}=\tfrac{1}{\Lambda}c^{I}\vev{k^{Ju}f_{IJ}^{K}\mu_{K}}+c^{I}\vev{f_{JI}^{K}h^{J}k_{K}^{u}}\ .
\end{equation}
Using again that $\vev{\mu_{I}}=\Lambda \vev{h_{I}}$ and applying \eqref{eq:structureconstants}, this yields
\begin{equation}
\tfrac{1}{\Lambda}c^{I}\vev{k^{Ju}f_{IJ}^{K}\mu_{K}}+c^{I}\vev{f_{JI}^{K}h^{J}k_{K}^{u}}=(f_{JI}^{K}+f_{IJ}^{K})c^{I}\vev{h^{J}k_{K}^{u}}= 0\ .
\end{equation}
Thus the Goldstone directions $\delta q^{u}=c^{I}\vev{k_{I}^{u}}$ leave the vacuum conditions \eqref{eq:backgroundmomentmaps} invariant and hence $\cG \subset \cD$.

Let us now consider the moduli space $\cM = \cD / \cG$ and show that
$J^{3}(\cM)=\cM$, i.e.\ $J^{3}$ restricts to an almost complex
structure on $\cM$. Concretely we show that the defining equations for the moduli space, \eqref{eq:12} and \eqref{eq:Killing1}, are invariant under $J^{3}$. For equations (\ref{eq:12}) this follows from the fact that $J^{3}$ interchanges the two equations. This can be seen by substituting $\delta q'^{u} = (J^{3})^{u}_{v}\delta q^{v}$ and using that $J^{1}J^{2}=J^{3}$ on a quaternionic K\"ahler manifold. 

Turning to \eqref{eq:Killing1}, we note that since only
$\vev{\mu_{I}^{3}}\neq 0$ the covariant derivative
\eqref{eq:Jinvariance} of the Killing vectors $k_{I}^{u}$  commutes
with $J^{3}$ in the vacuum, i.e.
\begin{equation}
 \vev{\nabla_{u}k^{I}_{w}(J^{n})_{v}^{w}-(J^{n})_{u}^{w}\nabla_{w}k^{I}_{v}}=2\epsilon^{npq}\vev{\omegaC^{p}_{uv}\mu^{Iq}} = 0\ .
\end{equation}
This implies that the second term in \eqref{eq:Killing1} is invariant
under $J^{3}$ and we need to show that this also holds for the first
term. In fact, we will show in the following
that this term vanishes on the moduli space and is only 
nonzero for Goldstone directions.

Let us first note that in general
$\rk{\vev{k_{I}^{u}\omegaC_{vw}^{3}k^{wI}}}\leq\rk{\vev{k_{I}^{u}}}=n_{G}$.
However, 
$\vev{k_{I}^{u}\omegaC_{vw}^{3}k^{wI}k_{J}^{v}} \neq 0$ (as we
saw in \eqref{GBint}) implies that the rank of the two matrices has
to coincide. This in turn says that the first term in
\eqref{eq:Killing1}
can only be nonzero in the Goldstone directions and thus has to
vanish
for the directions spanning $\cM$. Thus the whole equation \eqref{eq:Killing1} is $J^{3}$-invariant on
$\cM$. 
Therefore we have an almost complex structure
$\tilde{J}:=J^{3}\vert_{\cM}$ and a compatible metric
$\tilde{G}:=G\vert_{\cM}$ on $\cM$. Thus $(\cM, \tilde{G}, \tilde{J})$ is an almost hermitian submanifold  of the quaternionic K\"ahler manifold $(\cT_{H}, G, Q)$. 

In the following we want to use theorem 1.12 of \cite{Alekseevsky:2001om}: an almost Hermitian submanifold $(M, G, J)$ of a quaternionic K\"ahler manifold $(\tilde{M}, \tilde{G}, Q)$ is K\"ahler if and only if it is totally complex, i.e.\ if there exists a section $I$ of $Q$ that anticommutes with $J$ and satisfies
\begin{equation}
I(T_{p}M) \perp T_{p}M \quad \forall p\in M\ .
\end{equation}
In particular, this condition is satisfied if the associated fundamental two-form $\omegaC_{uw}=G_{uw}I_{v}^{w}$ on $M$ vanishes.

Now let us show that the moduli space $\cM$ actually is totally
complex and hence K\"ahler. To do so, we use \eqref{eq:covderkilling}
and \eqref{SLdef} to 
note that in the vacuum
\eqref{eq:momentmapsvacuum} 
$ \vev{\omegaC^{3}_{uv}}$ is given by 
\begin{equation}\label{eq:omega3}
 \vev{\omegaC^{3}_{uv}}=\tfrac{2}{\Lambda}\vev{h^{I}\nabla_{u}k_{Iv}-L_{uv}}\ .
\end{equation}
We just argued that 
$\vev{k_{I}^{u}\omegaC_{vw}^{3}k^{wI}}$ vanishes on $\cM$
and thus \eqref{eq:Killing1} projected onto $\cM$ also implies
\begin{equation}\label{eq:nablaG}
 \vev{h^{I} \nabla_{u}k_{vI}}\vert_{\cM} = 0 \ .
\end{equation}
Since $\vev{\omegaC^{1}_{uv}}=-\vev{\omegaC^{3}_{uw}(J^{2})_{v}^{w}}$, we can multiply \eqref{eq:omega3} with $-(J^{2})^{w}_{v}$ from the right and obtain
\begin{equation}
\vev{\omegaC^{1}_{uv}}\vert_{\cM} =
\tfrac{2}{\Lambda}\vev{S^{2}_{uv}-h^{I}
  \nabla_{u}k_{wI}(J^{2})_{v}^{w}}\vert_{\cM}=0\ ,
\end{equation}
where in the first step we used \eqref{SLdef}.
This expression  vanishes due to 
\eqref{eq:nablaG} and the fact that $S^{2}_{uv}$ is symmetric while
$\omegaC^{1}_{uv}$ is antisymmetric.
Thus $\cM$ is totally complex and in particular $(\cM, \tilde{G}, \tilde{J})$ is a K\"ahler submanifold. 

As proved in \cite{Alekseevsky:2001om} a K\"ahler submanifold can
have at most half the dimension of the ambient quaternionic K\"ahler
manifold, i.e.\ $\text{dim}(\cM) \leq 2n_{H}$.\footnote{Applying the same method as in $d=4$, $\cN=2$ this can be checked explicitly \cite{deAlwis:2013jaa}.}
Note that in the case of an unbroken gauge group we have $\cG = \{\emptyset\}$ and thus $\cD=\cM$. This is the case of maximal dimension of the moduli space. If the gauge group is now spontaneously broken then additional scalars are fixed by \eqref{eq:12}. Since $\cM$ is $J^{3}$-invariant, every $\delta q^{u} \in \cM$ can be written as $\delta q^{u} = (J^{3})_{v}^{u}\delta q'^{v}$ for some $\delta q'^{u}\in \cM$. Combined with the fact that $J^{1}J^{2}=J^{3}$ this implies that the two conditions in \eqref{eq:12} are equivalent on $\cM$. Furthermore we have $\rk{\vev{h^{I}_{i}\omegaC_{uv}^{1}k_{I}^{v}}}=\rk{\vev{k_{u}^{I}}}=n_{G}$ and thus $n_{G}$ scalars are fixed by \eqref{eq:12}. In conclusion, we altogether have
\begin{equation}
\text{dim}(\cM)=\text{dim}(\cD)-\text{dim}(\cG)\leq (2n_{H}-n_{G})-n_{G}\ ,
\end{equation}
so the moduli space has at most real dimension $2n_{H}-2n_{G}$.


\section*{Acknowledgments}
This work was supported by the German Science Foundation (DFG) under
the Collaborative Research Center (SFB) 676 ``Particles, Strings and the Early
Universe'', the Research Training Group (RTG) 1670 ``Mathematics
inspired by String Theory and Quantum Field Theory'' and the Joachim-Herz Stiftung.

We have benefited from conversations and correspondence with David Ciupke, Peter-Simon Dieterich, Malte Dyckmanns, Jonathan Fisher, Severin L\"ust, Stefan Vandoren and Owen Vaughan.




\newpage 

\providecommand{\href}[2]{#2}\begingroup\raggedright\endgroup

\end{document}